\documentclass[runningheads]{llncs}
\usepackage{graphicx}
\usepackage{tabularx}
\usepackage{hhline}
\usepackage{todonotes}
\usepackage{xcolor}
 \usepackage{multirow}

\begin{document}

\title{Monitoring Achilles tendon healing progress in ultrasound imaging with convolutional neural networks}

\author{
Piotr Woznicki\inst{1, 2} \and
Przemyslaw Przybyszewski\inst{3, 2} \and
Norbert Kapinski\inst{2} \and
Jakub Zielinski\inst{2} \and
Beata Ciszkowska-Lyson\inst{4} \and
Bartosz A. Borucki\inst{2} \and
Tomasz Trzcinski\inst{5, 6} \and
Krzysztof S. Nowinski\inst{2}}

\institute{Medical University of Warsaw \and
University of Warsaw \and
SGH Warsaw School of Economics \and
Carolina Medical Center \and
Warsaw University of Technology \and Tooploox}
\maketitle

\begin{abstract}
Achilles tendon rupture is a debilitating injury, which is typically treated with surgical repair and long-term rehabilitation. The recovery, however, is protracted and often incomplete. Diagnosis, as well as healing progress assessment, are largely based on ultrasound and magnetic resonance imaging. In this paper, we propose an automatic method based on deep learning for analysis of Achilles tendon condition and estimation of its healing progress on ultrasound images. We develop custom convolutional neural networks for classification and regression on healing score and feature extraction. Our models are trained and validated on an acquired dataset of over 250.000 sagittal and over 450.000 axial ultrasound slices. The obtained estimates show high correlation with the assessment of expert radiologists, with respect to all key parameters describing healing progress. We also observe that parameters associated with i.a. intratendinous healing processes are better modeled with sagittal slices. We prove that ultrasound imaging is quantitatively useful for clinical assessment of Achilles tendon healing process and should be viewed as complementary to magnetic resonance imaging.
\end{abstract}
\section{Introduction}
The Achilles tendon is the largest and strongest tendon in the human body. However, it is one of the most frequently injured tendons, especially among middle-aged people who participate in recreational sports. The incidence of Achilles tendon ruptures has been increasing over the last years~\cite{ZHOU20181191}. Usually, the diagnosis of an acute rupture is based on detailed musculoskeletal examinations and comprehensive medical history. Ultrasonography (US) and Magnetic Resonance Imaging (MRI) are routinely used for confirming the clinical diagnosis. 

The surgical treatment of acute Achilles tendon rupture has been shown to reduce the risk of re-rupture, but it might also lead to a higher complication rate~\cite{ZHOU20181191}. Furthermore, recent studies show that early functional rehabilitation could also stimulate tendon healing. For the above reasons, regular evaluation of the early tendon healing process is needed to establish patient prognosis and plan further treatment. The US findings correlate with several healing parameters, including cross-sectional area, tendon length or intratendinous morphology and are considered a safe and convenient method of assessing the healing progress~\cite{pmid30135861}. However, some studies have found only a moderate correlation of US findings with clinical assessment of Achilles tendinopathy and clinical outcomes~\cite{Khan149}.

Quantitative methods based on deep learning are well-suited for modelling the complex relationships between medical images and their interpretation. Recently, approaches using convolutional neural networks (CNNs) have outperformed traditional image analysis methods and proved their usefulness in the analysis of the Achilles tendon MRI scans~\cite{Kapinski2018}. 

In this study, we present a method for the automatic evaluation of the healing process of reconstructed Achilles tendon based on CNNs. We extend the approach proposed in~\cite{Kapinski2018} to US images in the axial and the sagittal plane and develop a novel method for healing phase estimation. To our knowledge, there are no other approaches in the literature to quantitatively asses the process of tendon healing through automated analyses of MRI and US imaging. Within this paper we also show that the method applied to MRI cannot by directly transferred to US data, which might result from problematic interpretation of the US images. 

More precisely, we first train and evaluate neural networks for the task of binary classification of a single ultrasound slice as healthy or injured. We then present our approaches to modelling the healing progress with respect to 6 key healing parameters. We analyse the applicability of the method using outputs of a pre-trained network with a linear classifier on the PCA-reduced space of the features to assess the progress with the US data. We find that this method fails to learn the accurate representation of the healing phase, therefore we propose an end-to-end CNN performing regression on healing parameters as a new, alternative approach. We further discuss the meaningfulness of the results for US and compare them with MRI results, to finally determine the clinical usefulness of used modalities and applicability of automatic methods for healing assessment.

\section{Methods}

In this section we describe our method based on the Convolutional Neural Networks. CNNs are discriminative deep architectures, able to extract high-level spatial and configuration information from an image, thus making them suitable for classification of 2D US imaging.

We use models with weights pretrained on ImageNet and train them to explicitly model radiologist assessments. 
To this end, we modify the architecture of the top dense layer of the CNN in such a way that the output layer performs linear regression on the high-level features from the penultimate layer. 
For initial tests we use three models of various complexity to eventually select Inception-v3~\cite{Szegedy15} architecture as a base for our final solution. These experiments are described as the supervised approach.
We then exploit the latent representation and reduce the dimensionality, which makes it possible to obtain a single-number summary of the tendon condition on one US examination. We refer to it as semi-supervised approach. 
In general, our approach leverages the ability of neural networks to approximate non-linear mappings directly and implicitly accounts for the intermediate feature representations. It maps the images to the tendon healing scores for the different protocols and clinical parameters. We train separate models for both US planes and for all of the ground-truth parameters described in the next subsection. 

\begin{figure*}[h]
\vspace{-0.5cm}
\centering
\includegraphics[width=0.95\textwidth]{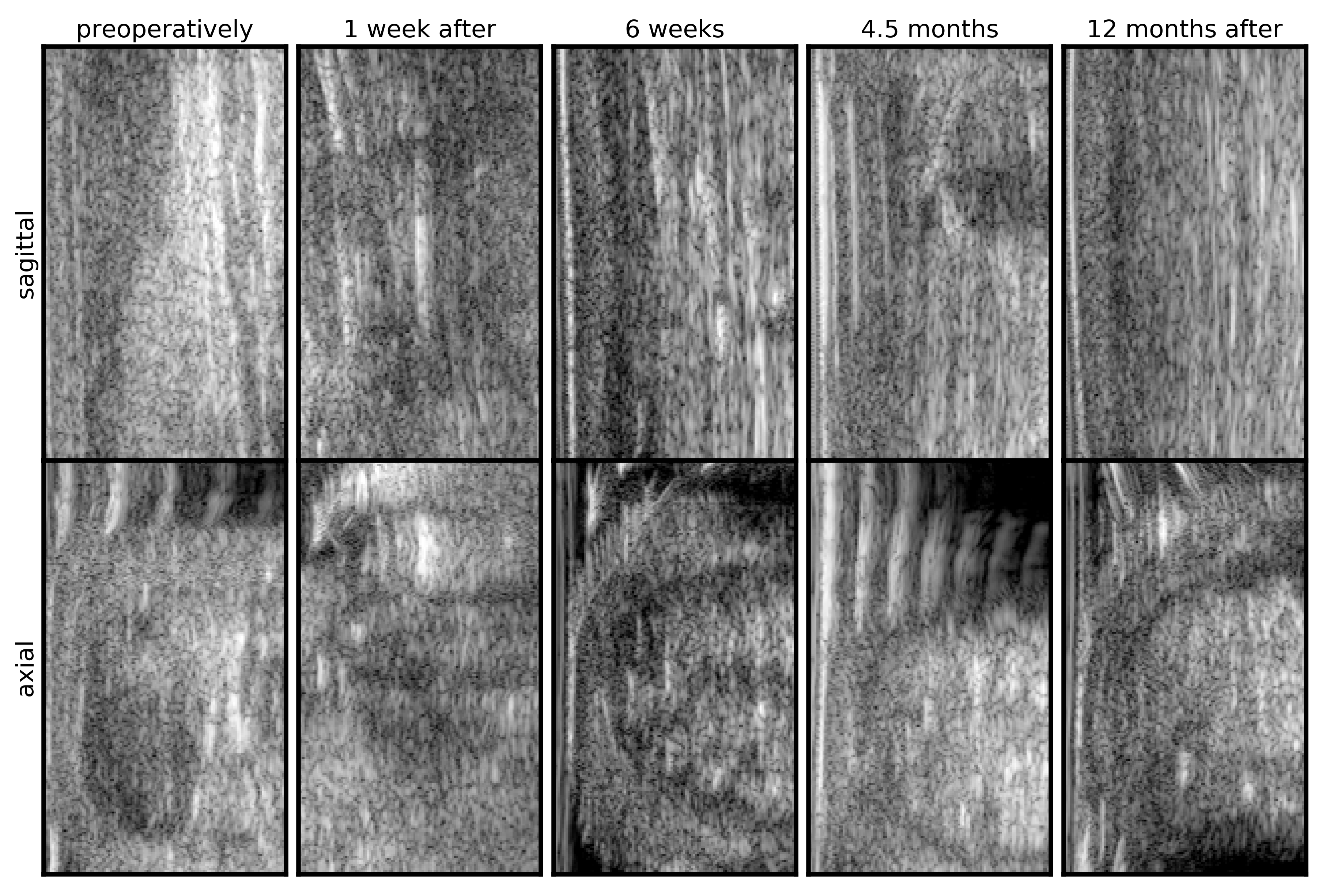}
\caption{Achilles healing process for a chosen patient on US imaging. 
On the sagittal images one can observe the gradual recovery of the fibrillary pattern of tendon fibres with hyperechoic bands.
In the axial plane the typical change is the widening of the Achilles tendon and the loss of the hypoechoic fluid collection surrounding the tendon. 
The images also exemplify certain artifacts typical for ultrasound, including reverberation, refraction and acoustic shadowing.}
\label{fig:usg}
\vspace{-0.5cm}
\end{figure*}

\subsection{Healing progress scoring} 
Our ground-truth is a survey that has been devised by expert radiologists, in order to quantitatively characterize their subjective assessment of Achilles tendon healing progress based on MRI and US. The survey evaluates the anatomy, metabolic activity and general functionality of the tendon. The following 6 parameters describing the tendon healing process were proposed~\cite{Kapinski2018}:
\begin{enumerate}
\item Structural changes within the tendon (SCT)
\item Tendon thickening (TT)
\item Sharpness of the tendon edges (STE)
\item Tendon edema (TE)
\item Tendon uniformity (TU)
\item Tissue edema (TisE)
\end{enumerate}
Each parameter is evaluated on a 7-point scale, where 1 corresponds to healthy and 7 to severely injured tendon. We use the scores as ground-truth labels in the training process. Our image dataset is presented in the next subsection.

\subsection{Dataset}
The original ultrasound dataset includes 49 patients with acute Achilles tendon rupture, all of whom underwent repair surgery and were closely monitored thereafter. The age of patients ranged from 18 to 50 years with a mean age of 36 years. The ultrasound examination was performed at 10 respective intervals: preoperatively, 1 week, 3, 6, 9, 12 weeks after, 4.5, 6, 9 and 12 months after the reconstruction. Additionally, 18 healthy volunteers have been scanned once. For all the examinations a GE 3D high-resolution Voluson E8 Expert ultrasound machine has been used with linear probes 5--18 MHz. The total dataset consists of 565 3D US exams but in this work, we focus on 2D scans only. Clinically, sagittal and axial scanning planes are used interchangeably by rotating the transducer, so we conduct the experiments separately for both. Considering the 2D slices, the final dataset includes 253,639 sagittal scans, 245,366 from patients with ruptured tendon and 8,273 from healthy patients. Alternatively, it consists of 467,548 axial scans, 450,816 injured and 16,732 healthy. The healing progression for an exemplary patient is shown in Fig. \ref{fig:usg}.
Though a detailed analysis can be done only by a trained medical professional, one can observe that the filamentous structures are more visible on the sagittal cross-sections while axial slices present in more details the tissue surrounding, edema and internal tendon pattern.

\begin{table}[h]
\scriptsize
\centering
\setlength{\tabcolsep}{3pt}
\setlength\extrarowheight{2pt}
\caption{Five-fold cross-validation results for the balanced dataset of 2D US scans.}
\label{tab1}
\begin{tabular}{l||c|c|c||c|c|c}
& \multicolumn{3}{c}{\footnotesize{\textbf{sagittal}}} & \multicolumn{3}{c}{\footnotesize{\textbf{axial}}} \\
Architecture & \textbf{Accuracy} & Precision & Recall & \textbf{Accuracy} & Precision & Recall \\ \hline
AlexNet & 0.846$\pm$00.087 & 0.92$\pm$00.08 & 0.78$\pm$00.11 & 0.843$\pm$00.075 & 0.93$\pm$00.06 & 0.73$\pm$00.11  \\ \hline
Inception-v3 & \textbf{0.916}$\pm$00.049 & 0.97$\pm$00.04 & 0.90$\pm$00.06 & 0.901$\pm$00.052 & 0.95$\pm$00.3 & 0.87$\pm$00.07 \\ \hline
ResNet50 & 0.907$\pm$00.039 & 0.96$\pm$00.05 & 0.89$\pm$00.08 & \textbf{0.912}$\pm$00.046 & 0.95$\pm$00.04 & 0.88$\pm$00.06 \\ \hline
\end{tabular}
\end{table}

\section{Experiments}

\subsection{Binary classification}
We train three network architectures: AlexNet~\cite{AlexNet}, Inception-v3 and ResNet50~\cite{HeZRS15} independently on sagittal and axial slices for the task of binary classification of the tendon on a 2D US scan as healthy or injured. The injured class is represented by all the exams of ruptured Achilles tendon performed preoperatively or 1 week after surgery. In order to balance the two classes we use mirroring on the healthy slices and we subsample injured patients for every training epoch. 

\begin{figure*}[h]
\centering
\includegraphics[width=1.0\textwidth]{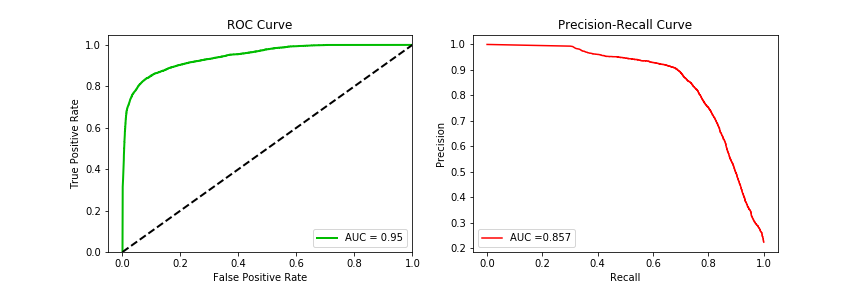}
\caption{ROC and Precision-Recall Curves for the Inception-v3 on sagittal US images.}
\label{fig:ROC_PR}
\end{figure*}

The accuracy is assessed in 5-fold cross-validation (Tab. \ref{tab1}). ROC and Precision-Recall Curves of the best performing model in terms of highest accuracy (Inception-v3 on sagittal slices) are presented in Fig. \ref{fig:ROC_PR}. For both Inception-v3 and ResNet50 we obtained an accuracy of over $90\%$ on both sagittal and axial scans, which proves that a CNN can be successfully trained on ultrasound data to differentiate between healthy and injured state.

We also experiment with the region of interest (ROI) segmentation as a preprocessing step for sagittal scans, applying {\it Active Contours Without Edges}~\cite{Chan2001}, which is widely used in the medical field. We hypothesize that focusing exclusively on the tendon region might reduce the noise and artifacts inherently present in US imaging. However, the experiments show lower accuracy with ROI segmentation cropping as compared to non-cropped images, which suggests that the tissues surrounding the Achilles tendon contribute relevant information to the classification.

\subsection{Healing progress estimation:} 
\subsubsection{Semi-supervised approach} 
The neural networks trained for binary classification are used as feature extractors for the task of computing the healing progress score. Principal Component Analysis (PCA) is applied on the feature space to reduce its dimensionality and the first principal component is considered as a representative score for the 2D US scan. For every examination, the aggregate score is calculated as a truncated mean of all 2D scan scores within a single study. 

\begin{table}[h]
\scriptsize
\setlength{\tabcolsep}{3pt}
\centering
\caption{5CV results for the tendon healing progress using end-to-end approach}
\label{tab2}
\begin{tabular}{lc||c|c|c|c|c|c}
& & \multicolumn{6}{c}{\footnotesize{\textbf{sagittal}}} \\
\textbf{Network} & & \textbf{SCT} & \textbf{TT} & \textbf{STE} & \textbf{TE} & \textbf{TU} & \textbf{TisE} \\ \hline
\multirow{3}{*}{AlexNet} & \textbf{MAE} & 0.96$\pm$0.41 & 0.80$\pm$0.27 & 0.82$\pm$0.29 & 0.95$\pm$0.41 & 0.87$\pm$0.39 & 1.08$\pm$0.47  \\
                   & MAX-AE & 1.75 & 1.32 & 1.87 & 1.35 & 1.61 & 2.03 \\ 
                   & Corr & 0.53$\pm${0.47} & 0.69$\pm${0.38} & 0.11$\pm${0.32} & 0.68$\pm${0.44} & 0.31$\pm${0.51} & 0.22$\pm${0.53} \\ \hline
\multirow{3}{*}{Inception-v3} & \textbf{MAE} & \textbf{0.88}$\pm$0.35 & \textbf{0.67}$\pm$0.23 & \textbf{0.80}$\pm$0.31 & \textbf{0.82}$\pm$0.23 & \textbf{0.84}$\pm$0.32 & \textbf{0.93}$\pm$0.29  \\
                   & MAX-AE & 1.69 & 1.32 & 1.69 & 1.31 & 1.58 & 1.64 \\ 
                   & Corr & 0.83$\pm${0.44} & 0.71$\pm${0.40} & 0.19$\pm${0.34} & 0.64$\pm${0.47} & 0.56$\pm${0.40} & 0.71$\pm${0.40} \\ \hline
\multirow{3}{*}{ResNet50} & \textbf{MAE} & \textbf{0.89}$\pm$0.12 & \textbf{0.74}$\pm$0.15 & \textbf{0.83}$\pm$0.22 & \textbf{0.81}$\pm$0.31 & \textbf{0.92}$\pm$0.31 & \textbf{0.99}$\pm$0.32 \\
                   & MAX-AE & 1.53 & 1.22 & 1.64 & 1.43 & 1.67 & 1.71 \\
                   & Corr & 0.62$\pm${0.31} & 0.38$\pm${0.51} & 0.23$\pm${0.41} & 0.62$\pm${0.51} & 0.12$\pm${0.43} & 0.43$\pm${0.50} \\ \hhline{|=|=|=|=|=|=|=|=|}
& & \multicolumn{6}{c}{\footnotesize{\textbf{axial}}} \\
& & \textbf{SCT} & \textbf{TT} & \textbf{STE} & \textbf{TE} & \textbf{TU} & \textbf{TisE}\\ \hline
\multirow{3}{*}{AlexNet} & \textbf{MAE} & 0.98$\pm$0.39 & 0.83$\pm$0.33 & 0.82$\pm$0.35 & 0.94$\pm$0.52 & 0.95$\pm$0.50 & 0.86$\pm$0.28  \\
                   & MAX-AE & 1.79 & 1.36 & 1.86 & 1.41 & 1.61 & 1.59 \\ 
                   & Corr & 0.45$\pm${0.33} & 0.62$\pm${0.39} & 0.20$\pm${0.45} & 0.60$\pm${0.47} & 0.03$\pm${0.42} & 0.59$\pm${0.40} \\ \hline
\multirow{3}{*}{Inception-v3} & \textbf{MAE} & \textbf{1.03}$\pm$0.46 & \textbf{0.70}$\pm$0.24 & \textbf{0.76}$\pm$0.26 & \textbf{0.86}$\pm$0.19 & \textbf{0.87}$\pm$0.32 & \textbf{0.85}$\pm$0.25  \\
                   & MAX-AE & 2.52 & 1.45 & 1.45 & 1.24 & 1.67 & 1.56 \\ 
                   & Corr & 0.77$\pm${0.47} & 0.69$\pm${0.40} & 0.22$\pm${0.37} & 0.65$\pm${0.41} & 0.55$\pm${0.44} & 0.72$\pm${0.41} \\ \hline
\multirow{3}{*}{ResNet50} & \textbf{MAE} & \textbf{1.05}$\pm$0.31 & \textbf{0.78}$\pm$0.33 & \textbf{0.80}$\pm$0.24 & \textbf{1.02}$\pm$0.25 & \textbf{0.87}$\pm$0.15 & \textbf{0.91}$\pm$0.28 \\
                   & MAX-AE & 1.98 & 1.51 & 1.59 & 1.63 & 1.45 & 1.57 \\
                   & Corr & 0.52$\pm${0.41} & 0.47$\pm${0.44} & 0.22$\pm${0.35} & 0.65$\pm${0.55} & 0.18$\pm${0.54} & 0.57$\pm${0.39} \\ \hhline{|=|=|=|=|=|=|=|=|}
\end{tabular}
\end{table}

Although this method was proven to work for MRI scans~\cite{Kapinski2018}, for ultrasound we observed a very weak correlation with actual healing parameters, which should be attributed to lower variance preserved by the first principal components and higher variance between scans from one examination. Therefore we do not present the results here. We believe that speckle noise, a random granular pattern produced mainly by multiplicative disturbances, as well as frequent artifacts are the main reasons for the weak performance of the tested method.

\subsubsection{Supervised approach} 
Healing scores are evaluated in 5-fold cross-validation using mean absolute error (MAE), maximal absolute error for a single exam (MAX-AE) and mean correlation, computed with the use of Fisher Z-Transformation (Tab. \ref{tab2}).

We observe a good correspondence between the estimated healing scores and the experts' assessment, with MAE ranging from $0.67$ to $1.08$, on a 7 point scale. For all the networks we notice a positive mean correlation of our method's output and healing parameters. Although the results are consistent between different networks, Inception-v3 usually achieves the best fit and the simplest network architecture, AlexNet, performs noticeably worse. Two healing parameters, SCT and TT are more accurately estimated on sagittal rather than axial US images and one parameter, TisE, vice versa.

\begin{figure*}[h]
\centering
\includegraphics[width=1.0\textwidth]{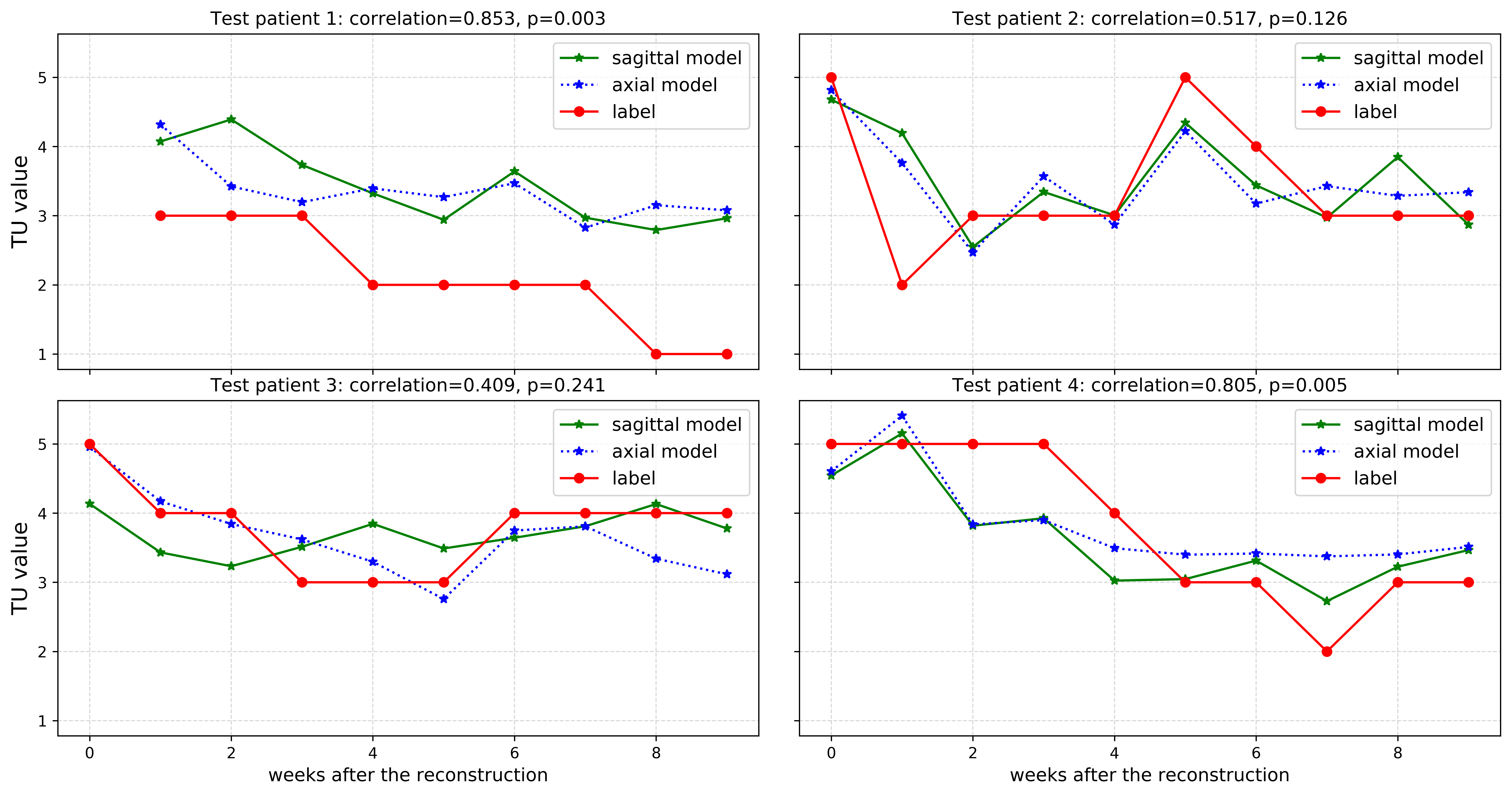}
\caption{Inception-v3 results for the TU parameter on test dataset (correlations refer to sagittal scans).}
\label{fig:Incv3_TU}
\end{figure*}

The final evaluation of the regression task has been done on a separate test set, consisting of 4 injured patients who underwent a full rehabilitation process, i.e. 40 studies in total (Tab. \ref{tab3}). For the best performing Inception-v3, we report MAE ranging from $0.53$ to $0.87$ and correlations in the range of $0.31$ to $0.80$.
The resulting healing progress for a selected parameter is compared with radiologist evaluation in Fig. \ref{fig:Incv3_TU}. In general, axial and sagittal models give similar results, which tend to correlate well with ground-truth labels.

\begin{table}[h]
\footnotesize
\setlength{\tabcolsep}{1pt}
\centering
\caption{Results for the tendon healing progress using end-to-end approach on the test dataset}
\label{tab3}
\begin{tabular}{lc||c|c|c|c|c|c}
& & \multicolumn{6}{c}{\normalsize{\textbf{sagittal}}} \\
\textbf{Network} & & \textbf{SCT} & \textbf{TT} & \textbf{STE} & \textbf{TE} & \textbf{TU} & \textbf{TisE} \\ \hline
\multirow{2}{*}{AlexNet} & \textbf{MAE} & 0.90$\pm$0.31 & 0.63$\pm$0.12 & 0.69$\pm$0.31 & 0.81$\pm$0.11 & 0.89$\pm$0.20 & 1.01$\pm$0.35 \\
                   & Corr & 0.55$\pm$0.15 & 0.70$\pm$0.24 & 0.22$\pm$0.49 & 0.61$\pm$0.28 & 0.12$\pm$0.43 & 0.28$\pm$0.27 \\ \hline
\multirow{2}{*}{Inception-v3} & \textbf{MAE} & \textbf{0.81}$\pm$0.38 & \textbf{0.63}$\pm$0.06 & \textbf{0.56}$\pm$0.18 & \textbf{0.85}$\pm$0.20 & \textbf{0.54}$\pm$0.04 & \textbf{0.87}$\pm$0.29 \\
                   & Corr & 0.80$\pm$0.39 & 0.77$\pm$0.28 & 0.31$\pm$0.33 & 0.52$\pm$0.36 & 0.69$\pm$0.34 & 0.62$\pm$0.52 \\ \hline
\multirow{2}{*}{ResNet50} & \textbf{MAE} & \textbf{0.88}$\pm$0.33 & \textbf{0.65}$\pm$0.15 & \textbf{0.66}$\pm$0.09 & \textbf{0.83}$\pm$0.25 & \textbf{0.75}$\pm$0.12 & \textbf{0.93}$\pm$0.22 \\
                   & Corr & 0.60$\pm$0.32 & 0.55$\pm$0.38 & 0.25$\pm$0.27 & 0.55$\pm$0.41 & 0.34$\pm$0.29 & 0.56$\pm$0.38 \\
                   \hhline{|=|=|=|=|=|=|=|=|}
& & \multicolumn{6}{c}{\normalsize{\textbf{axial}}} \\
& & \textbf{SCT} & \textbf{TT} & \textbf{STE} & \textbf{TE} & \textbf{TU} & \textbf{TisE}\\ \hline
\multirow{2}{*}{AlexNet}  & \textbf{MAE} & 1.12$\pm$0.36 & 0.81$\pm$0.29 & 0.58$\pm$0.12 & 0.87$\pm$0.19 & 0.70$\pm$0.24 & 0.85$\pm$0.25 \\
                   & Corr & 0.46$\pm$0.50 & 0.54$\pm$0.32 & 0.26$\pm$0.41 & 0.38$\pm$0.38 & 0.12$\pm$0.34 & 0.70$\pm$0.31 \\ \hline
\multirow{2}{*}{Inception-v3} & \textbf{MAE} & \textbf{0.84}$\pm$0.54 & \textbf{0.75}$\pm$1.45 & \textbf{0.58}$\pm$0.10 & \textbf{0.83}$\pm$0.10 & \textbf{0.53}$\pm$0.16 & \textbf{0.83}$\pm$0.30 \\
                   & Corr & 0.69$\pm$0.49 & 0.68$\pm$0.41 & 0.45$\pm$0.15 & 0.51$\pm$0.42 & 0.66$\pm$0.16 & 0.68$\pm$0.39 \\ \hline
\multirow{2}{*}{ResNet50} & \textbf{MAE} & \textbf{0.92}$\pm$0.37 & \textbf{0.76}$\pm$0.32 & \textbf{0.68}$\pm$0.08 & \textbf{0.81}$\pm$0.17 & \textbf{0.65}$\pm$0.20 & \textbf{0.94}$\pm$0.11 \\
                   & Corr & 0.55$\pm$0.41 & 0.57$\pm$0.38 & 0.35$\pm$0.31 & 0.44$\pm$0.39 & 0.39$\pm$0.35 & 0.61$\pm$0.33 \\ \hhline{|=|=|=|=|=|=|=|=|}
\end{tabular}
\end{table}

\section{Discussion}
We show that a neural network learns to extract features from the US images which strongly correlate with the healing progress score assigned by expert radiologists. Out of the three healing parameters: tendon uniformity (TU), structural changes (SCT) and tendon thickening (TT), which correspond to morphological changes within the Achilles tendon and are typically evaluated in the longitudinal axis, SCT and TT are better modeled by the sagittal ultrasound, while TU still retains MAE of $<1$ point. On the other hand, sharpness of the tendon edges (STE), tendon edema (TE) and tissue edema (TisE) are typically evaluated on axial slices and for STE and TisE, all our networks achieve lower MAE and higher mean correlation when trained in the axial plane. 

In comparison with the results from~\cite{Kapinski2018}, we notice that a convolutional neural network is able to achieve a better accuracy of binary classification on MRI data rather than US data (99.83\% vs. 91.6\% for the best respective models). Furthermore, a high
correlation of automated method output with the ground truth in terms of three parameters: TE, TisE and STE has been reported for MRI scans. MR-acquired stacks of axial images of the Achilles tendon have a major limitation in the form of lower spatial resolution along the longitudinal axis, which is determined by the slice selection pulse. Because of this spatial anisotropy, they are not suitable for assessing healing parameters, which rely on the intratendinous processes or the alignment of fibrous bands. 

The results suggest that features extracted by deep learning models from MR and US imaging focus on different qualities of the rehabilitation process. This indicates that ultrasound should be viewed as an imaging method that complements MRI rather than one that competes with MRI in the evaluation of musculoskeletal abnormalities. It should be noted, however, that the previous work on MRI was validated on a smaller dataset and did not apply the supervised end-to-end approach, which limits us to an indirect qualitative comparison.

\section{Conclusions}
In this paper, we proposed deep learning models that achieve high performance in clinical classification and healing phase estimation of ruptured Achilles tendon. We have compared two approaches to modelling tendon rehabilitation progress and shown that the supervised method is superior to the semi-supervised method. Currently, monitoring the healing process requires a radiologist to analyze US and MRI data and subjectively evaluate the condition of the tendon.

As suggested in~\cite{pmid30135861}, tendon morphology may be the more robust measure to gauge patient healing progress over time compared to mechanical properties of the tendon. Therefore, we believe that a model which accurately estimates healing parameters from standardized images may be useful in clinical practice. 

Future studies are needed to improve the generalizability of deep learning models for medical imaging in musculoskeletal disorders and to determine the effect of model assistance in the clinical setting.

\bibliographystyle{ieeetr}
\bibliography{biblio}

\end{document}